\documentclass[a4paper,oneside,12pt,reqno]{amsart}	
\usepackage{amsmath,amssymb,latexsym,bm}
\usepackage{fancyhdr}

\begin{document}

\title{Gravitational stability of finite massive bodies}

\author{A.W.~Zaharow}

\address{Ufa state oil technic university, Ufa, Russia}
\email{zaharow@gmail.com}

\date{\today}

\numberwithin{equation}{section}

\begin{abstract}
Jeans instability of finite massive bodies at hydrostatic equilibrium
is studied. Differential equation governing the evolution of 
infinitesimal disturbances is derived. We take into account radial
inhomogeneity of mass density and other fluid parameters at 
the equilibrium state. Dispersion relation and a simple analytical
formula, generalizing the Jeans criterion of instability, are derived.
\end{abstract}

\maketitle
\thispagestyle{empty}

\section{introduction}\label{S:intro} 
It is well known (See, for instance \cite{jZ81} - \cite{aS04}) that the 
Jeans' instability forms the basis of our understanding of gravitational
condensation.
In particular, Jeans' mass criterion is invoked in astrophysical theories
of the formation of stars, gaseous clouds, etc. Usually gravitational 
instability is analysed in terms of the Jeans' wavelength \cite{mK99}
\begin{equation} \label{eq:jwl}
\lambda_J = \sqrt{\frac{\pi c_s^2}{G \rho_0}} \,,
\end{equation}
or, equivalently, in terms of Jeans' mass $M_J \sim \rho_0 \lambda_J^3$.
In this formula $G$ is the gravitational constant, $\rho_0$ is the
unperturbed mass density and $c_s$ is the adiabatic sound speed.
As is now widely known, perturbations in homogeneous fluid with mass greater
than a critical value $M_J$ may grow producing gravitationally bounded
structures. In the process of their evolution this structures can
achieve states of hydrodynamic equilibrium like stars polytropes or 
gas clouds when pressure gradient equals gravitational force.

In this paper we investigate hydrodynamic equilibrium and stability
of finite self-gravitating fluid mass with inhomogeneous distribution
of mass density, pressure and temperature along the radius.
In linear approach we get Schr\"odiger-like \cite{lL74} equation 
wich eigenvalues and eigenfunctions give us increments and
profiles of disturbances.

\section{basic formalism}\label{S:base} 
Consider spherically symmetric fluid body with radius $R$ and mass $M$.
We assume that the system is non-rotating and non-expanding. It can be
star, gaseous cloud, etc. The evolution of a self-gravitating fluid is
described by the conservation equations for mass, momentum and specific
entropy, coupled with the Poisson equation
\begin{equation} \label{eq:cont}
\frac{\partial \rho}{\partial t} + \mathbf{\nabla} \rho \mathbf{v} = 0
\end{equation}
\begin{equation} \label{eq:eul}
\rho \frac{\partial \mathbf{v}}{\partial t} + 
\rho (\mathbf{v}\mathbf{\nabla}) \mathbf{v} = 
- \mathbf{\nabla} p - \rho \mathbf{\nabla} \varphi
\end{equation}
\begin{equation} \label{eq:ent}
\frac{\partial s}{\partial t} + 
(\mathbf{v}\mathbf{\nabla}) s = 0
\end{equation}
\begin{equation} \label{eq:newt}
\nabla^2 \varphi = 4\pi G \rho \,,
\end{equation}
where $\mathbf{v}$ is the velocity, $p$ is the pressure, $\rho$ is the
mass density, $s$ is the specific entropy and $\varphi$ is the gravitational
potential. The linearization procedure has clearly invoked that local 
state variables deviate from their equilibrium values through linear 
fluctuation, namely
\begin{equation} \label{eq:lin}
\begin{split}
p =& p_0(r) + p_1(t,\mathbf{r})\,, \quad 
\rho = \rho_0(r)+\rho_1(t,\mathbf{r})\,, \\
s =& s_0(r) + s_1(t,\mathbf{r})\,,\quad 
\varphi = \varphi_0(r)+\varphi_1(t,\mathbf{r})\,.
\end{split}
\end{equation}
Velocity $\mathbf{v}$ itself is infinitesimal.
Substitution of (\ref{eq:lin}) in (\ref{eq:cont}) - (\ref{eq:newt})
constitutes equations for equilibrium state (\cite{jZ81})
\begin{equation} \label{eq:eul0}
-\frac{1}{\rho_0(r)}\mathbf{\nabla} p_0(r) - \mathbf{\nabla} \varphi_0(r)=0
\end{equation}
\begin{equation} \label{eq:newt0}
\nabla^2 \varphi_0(r) = 4\pi G \rho_0(r)
\end{equation}
and for perturbed parameters
\begin{equation} \label{eq:cont1}
\frac{\partial \rho_1}{\partial t} +
\mathbf{\nabla} \rho_0\mathbf{v} =0
\end{equation}
\begin{equation} \label{eq:eul1}
\rho_0\frac{\partial \mathbf{v}}{\partial t}=
-\mathbf{\nabla} p_1 - \rho_1\mathbf{\nabla} \varphi_0 -
\rho_0\mathbf{\nabla} \varphi_1
\end{equation}
\begin{equation} \label{eq:ent1}
\frac{\partial s_1}{\partial t} +
\mathbf{v} \mathbf{\nabla} s_0 = 0
\end{equation}
\begin{equation} \label{eq:newt1}
\nabla^2 \varphi_1 = 4\pi G \rho_1 \,.
\end{equation}
These equations must be coupled with the proper equation of state.
To simplify the analysis and exclude buoyancy forces (\cite{lL88},\cite{lR02})
we take it to be adiabatic 
\begin{equation} \label{eq:pres}
p \sim \rho^{\gamma} \,, 
\end{equation}
where $\gamma$ is the adiabatic exponent. 
Thus we get no entropy desturbances
\begin{equation} \label{eq:ent2}
s(t,\mathbf{r}) = s_0(r) = \text{const}
\end{equation}
and pressure and mass density disturbances are bound together with adiabatic
equation
\begin{equation} \label{eq:pres1}
p_1 = c_s^2 \rho_1 \,.
\end{equation}
Equations (\ref{eq:eul0}), (\ref{eq:newt0}) and (\ref{eq:pres})
for equilibrium state are resolved with well known Emden functions 
(\cite{jZ81}) for polytropic model. For further references we summarize
here basic results of polytropic theory of ideal gas when polytrope
exponent $n = 1/(\gamma - 1)$. In this case radial dependancy of equilibrium
parameters is
\begin{equation} \label{eq:poly1}
\begin{split}
p_0(r) & =  p(0) \Theta_n^{n+1}(\xi) \\
\rho_0(r) & = \rho(0) \Theta_n^{n}(\xi) \\
c_s^2(r) & = c_s^2(0) \Theta_n(\xi) \,,
\end{split}       
\end{equation}       
where $\xi = r/R$ and $\Theta_n$ are the non-dimensional radius and
temperature respectively.  To satisfy equilibrium equations (\ref{eq:eul0}),
(\ref{eq:newt0}) and boundary conditions $\Theta(0) = 1$, $\Theta(1) = 0$
parameter 
\begin{equation} \label{eq:xi1}
\xi_n^2 = \frac{4 \pi G \rho(0) R^2}{n c_s^2(0)}
\end{equation}       
must have unique value for each $n$. For example for $n = 3/2$ it
equals to $\xi_{3/2} = 3.65$ and for $n = 5/2$  $\xi_{5/2} = 5.36$
(\cite{jZ81}).

Equations (\ref{eq:cont1}), (\ref{eq:eul1}), (\ref{eq:newt1}) and
(\ref{eq:pres1}) with
proper boundary conditions forms the basis of our treatment of stability.

Boundary conditions we discuss later. But now we transform perturbed 
equations into the more convenient form. Substituting (\ref{eq:newt1})
into (\ref{eq:cont1}) gives us
\begin{equation} \label{eq:cont2}
\mathbf{\nabla}\left( \frac{1}{4\pi G} \mathbf{\nabla}
\frac{\partial \varphi_1}{\partial t} +
\rho_0\mathbf{v} \right) =0 \,.
\end{equation}
Taking into account that $\text{div rot}\mathbf{\Psi} \equiv 0$ we
get from (\ref{eq:cont2})
\begin{equation} \label{eq:vel1}
\rho_0 \mathbf{v} = - \frac{1}{4\pi G} \mathbf{\nabla}
\frac{\partial \varphi_1}{\partial t} + 
\text{rot} \mathbf{\Psi} = 0 \,,
\end{equation}
where $\mathbf{\Psi}$ stands for vector-potential of the flow 
$\rho_0 \mathbf{v}$. As we can see from (\ref{eq:vel1}), vector-potential 
$\mathbf{\Psi}$ represents the ``axial part'' of desturbances and is
not derectly bound with gravitational potential. So we assume it to be zero.
Next, inserting (\ref{eq:vel1}) with $\mathbf{\Psi} = 0$ and (\ref{eq:pres1})
into (\ref{eq:eul1}), we obtain
\begin{equation} \label{eq:eul2}
- \frac{1}{4\pi G} \mathbf{\nabla}
\frac{\partial^{2} \varphi_1}{\partial t^2} + 
\rho_0\mathbf{\nabla} \varphi_1 =
-\mathbf{\nabla} c_s^2 \rho_1 - \rho_1\mathbf{\nabla} \varphi_0 \,. 
\end{equation}
A harmonic time dependence $\sim \exp(-i\omega t)$ of the perturbations can 
now be assumed in terms of constant complex frequency $\omega$ so that 
equation (\ref{eq:eul2}) after simple transformations becomes
\begin{equation} \label{eq:eul3}
\mathbf{\nabla} \varphi_1 =
-4\pi G \frac{\mathbf{\nabla} c_s^2 \rho_1 + \rho_1\mathbf{\nabla} \varphi_0}
{\omega^2 + 4\pi G \rho_0} \,. 
\end{equation}
Taking divergence of equation (\ref{eq:eul3}) and using (\ref{eq:newt1})
we finally  get equation for mass density disturbance $\rho_1$ only
\begin{equation} \label{eq:main1}
\mathbf{\nabla} \left( \frac{\mathbf{\nabla} c_s^2 \rho_1 + 
\rho_1\mathbf{\nabla} \varphi_0}{\omega^2 + 4\pi G \rho_0} \right)
+ \rho_1 = 0 \,. 
\end{equation}
Before discussing equation (\ref{eq:main1}) we make the last simplification
by introducing auxillary function
\begin{equation} \label{eq:chi}
\chi(r)= \int^r_0 \frac{dr}{c_s^2 (r)}\frac{d}{dr} \left( \varphi_0(r) + 
c_s^2 (r) \right)
\end{equation}
and replace $\rho_1$ with new function $u$
\begin{equation} \label{eq:rho1}
\rho_1= \exp(-\chi)u \,.
\end{equation}
It yields final equation for unknown function $u$
\begin{equation} \label{eq:main2}
\exp(\chi)\mathbf{\nabla}\left(
\frac{\exp(-\chi) c_s^2}{\omega^2 + 4\pi G \rho_0}
\mathbf{\nabla} u \right) + u = 0 \,.
\end{equation}

Equation (\ref{eq:main2}) represents a modification of Sturm-Liouville
eigenvalue problem and, together with suitable boundary conditions,
provides the eigenvalues and eigenfunctions for the perturbations.
Boundary conditions are summarized as follows.

On the free moving surface of the body pressure must be equal to zero
\begin{equation} \label{eq:marg1}
p_1(t,r=R) = 0\,.
\end{equation}
Boundary conditions are stated at unperturbed surface so as small deviations
of size leads to second order magnitude of perturbance.
If the local sound velocity at the body surface is not equal zero, 
Eq. (\ref{eq:pres1}), (\ref{eq:marg1}) constitutes that mass density is
equal to zero
\begin{equation} \label{eq:marg2}
\rho_1(t,r=R) = 0\,, \quad \text{if } \quad c_s(r=R) \neq 0 \,.
\end{equation}
But if, as it is in polytropic models, $c_s(r=R) = 0$ then boundary
condition  (\ref{eq:marg1}) is valid with arbitrary $\rho_1(r=R)$.
To find out baundary condition for mass density in this case we integrate
(\ref{eq:cont1}) on the unperturbed volume. Taking into account Gauss 
theorem we get 
\begin{equation} \label{eq:mas1}
\frac{d}{dt}\int \rho_1 \,dV + \int \rho_0 \mathbf{v}\, d \mathbf{S} = 0\,,
\end{equation}
where $d \mathbf{S}$ is body surface element.
For zero mass density $\rho_0$ at $r = R$ second item in (\ref{eq:mas1})
is equal to zero and mass conservation low (\ref{eq:mas1}) reduces to
\begin{equation} \label{eq:mas2}
\int \rho_1 \,dV  = 0\,, \quad \text{if } \quad c_s(r=R) = 0 \,.
\end{equation}

A simplification can now be introduced to handle the non-trivial
angular dependence in the perturbation (\ref{eq:main2}). It can
be rewritten after decomposing variables into spherical harmonics
$Y_{lm}$
\begin{equation} \label{eq:harm1}
u(\mathbf{r}) = \sum_{l,m} U_{lm}(r)Y_{lm}(\theta,\phi) \,.
\end{equation}
Usually perturbations with quantum number $l > 0 $ have small
increments so we take into account only spherically symmetric
perturbations with $l = 0 $. For this case main equation
in spherical system of coordinates is
\begin{equation} \label{eq:main3}
Lu=\frac{\exp(\chi)}{r^2}\frac{d}{dr}\left( r^2
\frac{\exp(-\chi) c_s^2}{\omega^2 + 4\pi G \rho_0} 
\frac{du}{dr} \right) + u = 0 \,.
\end{equation}

Operator $L$ has some important features. First of all we 
now prove that it may have only real eigenvalues $\omega^2$.
Let multiply (\ref{eq:main3}) with complex conjugate $u^*$
and function $r^2 \exp(-\chi)$ and integrate along radius.
We get
\begin{equation} \label{eq:int1}
-\int_0^R \frac{\exp(-\chi) c_s^2}{\omega^2 + 4\pi G \rho_0}
|\frac{du}{dr} |^2 \,r^2 dr + \int_0^R \exp(-\chi)|u|^2\,r^2 dr = 0.
\end{equation}
First item in (\ref{eq:int1}) is derived with integrating by parts and
using that  $c_s^2 u$ equals zero on body surface. Subtracting from
(\ref{eq:int1}) its complex conjugate we get after simple transformation
\begin{equation} \label{eq:int2}
(\omega^2-\omega^{2*})
\int_0^R \frac{\exp(-\chi) c_s^2}{\lvert\omega^2 + 
4\pi G \rho_0 \rvert^2}
|\frac{du}{dr}|^2\,r^2 dr = 0 .
\end{equation}
Obviously integral in equation (\ref{eq:int2}) is greater zero and 
imaginary part of $\omega^2$ must be zero: $\text{Im}~ \omega^2 = 0$.
So there are two types of oscillation modes: if $\omega^2 > 0$ eigenvalues
$\omega$ are real and introduces sound-like oscillations, but if
$\omega^2 < 0$ eigenvalues are pure imaginary and branch with plus
imaginary part introduces monotonously growth of perturbation.

Next, again multiplying equation (\ref{eq:main3}) with $r^2 \exp(-\chi)$
and integrating it along $r$ we get
\begin{equation} \label{eq:mas3}
\left( r^2
\frac{\exp(-\chi) c_s^2}{\omega^2 + 4\pi G \rho_0} 
\frac{du}{dr} \right) \bigg|_0^R + 
\int_0^R \exp(-\chi) u r^2 \, dr = 0 \,.
\end{equation}
If $c_s(R) = 0$ first item in (\ref{eq:mas3}) equals zero and equation
(\ref{eq:main3}) automatically conserve total mass 
(cf. \ref{eq:mas2}, \ref{eq:rho1})
\begin{equation} \label{eq:mas4}
\int_0^R \exp(-\chi) u r^2 \, dr = 0 \,.
\end{equation}

To find out conditions for existance of instability suppose that 
$\omega^2 < 0$ and mass density $\rho_0(r)$ and sound velosity $c_s(r)$
monotonously decrease from theirs maxima at $r = 0$ to zero at $r = R$.
Introduce also function of wave number
\begin{equation} \label{eq:k1}
k(r) = \sqrt{\omega^2 + 4\pi G \rho_0(r)}/c_s(r) =
\sqrt{4\pi G \rho_0(r) - \lvert\omega\rvert^2 }/c_s(r)\,.
\end{equation}
In this notation equation (\ref{eq:main3}) can be rewritten as
\begin{equation} \label{eq:main4}
\frac{\exp(\chi)}{r^2}\frac{d}{dr}\left( r^2
\frac{\exp(-\chi)}{k^2} 
\frac{du}{dr} \right) + u = 0 \,.
\end{equation}

Wave number $k(r)$ equals zero at the radius $r_0$ where
\begin{equation} \label{eq:k2}
4\pi G \rho_0(r_0) - \lvert\omega\rvert^2 = 0 \,.
\end{equation}
At this point equation (\ref{eq:main4}) has singularity and eigenfunction
$u$ must have zero derivative
\begin{equation} \label{eq:u2}
\frac{du}{dr}(r=r_0) = 0\,.
\end{equation}

Now we build up eigenfunctions of (\ref{eq:main4}) with method like
quasiclassic approach in quantum mechanics (\cite{lL74}). ``Sewing
condition'' at the point $r = r_0$ will give us dispersion relation
for unstable modes.

Quasiclassic eigenfunction in the region $r < r_0$ finite at $r = 0$ is
\begin{equation} \label{eq:ul}
u = \frac{\exp(\chi / 2)k^{1/2}}{r} \sin 
\left( \int_0^r k(r)\,dr \right) 
\quad \text{ $r \ll r_0$, $k^2>0$.}
\end{equation}
As we can see from (\ref{eq:ul}), in this region perturbation oscillate
with radius. But in the region $r > r_0$ it exponentially decrease
\begin{equation} \label{eq:ur}
u = \frac{C \exp(\chi / 2)|k|^{1/2}}{2r} \exp 
\left( - \int_{r_0}^r |k(r)|\,dr \right)
\quad \text{ $r \gg r_0$, $k^2<0$},
\end{equation}
where $C$ is constant.
We will get ``sewing condition'' from analitical continuation of
expression (\ref{eq:ur}) into region $r < r_0$ through up and down
halfplanes of the complex variable $r-r_0$ (\cite{lL74}). This
procedure yields
\begin{equation} \label{eq:sew2}
\frac{C \exp(\frac{\chi}{2})|k|^{1/2}}{2r}\exp \left(-\int_{r_0}^{r} 
|k(r)|\,dr \right) \longrightarrow 
\frac{C \exp(\frac{\chi}{2})k^{1/2}}{r} \sin \left( \int_{r}^{r_0} k(r)\,dr +
\frac{3\pi}{4} \right).
\end{equation}
Right item of (\ref{eq:sew2}) must be equal to right item of (\ref{eq:ul})
\begin{equation} \label{eq:sew3}
C \sin \left( \int_{r}^{r_0} k(r)\,dr + \frac{3\pi}{4} \right) = 
\sin \left( \int_{0}^{r} k(r)\,dr \right) \,.
\end{equation}
Writing integral at right part of (\ref{eq:sew3}) in the form  
\begin{equation} \label{eq:sew4}
\int_{0}^{r} k(r)\,dr = \int_{0}^{r_0} k(r)\,dr - 
\int_{r}^{r_0} k(r)\,dr \,
\end{equation}
we get equality condition
\begin{equation} \label{eq:disp1}
\begin{split}
\int_{0}^{r_0} k(r)\,dr &= \frac{\pi}{4} + n \pi \\
n &=0,1,2, \ldots \\
C &= (-1)^n \,.
\end{split}
\end{equation}

Expressions (\ref{eq:disp1}) define discrete set of unstable modes
with encrements $\omega_n $ ($\omega_n^2 < 0$) and eigenfunctions
of type (\ref{eq:ul}), (\ref{eq:ur}). Eigenfunction $u_n$ has strictly
$n$ zero nodes at the interval $0<r<R$. But the mode with $n=0$ can not
exist as it does not change sign and can not obey the mass conservation
low (\ref{eq:mas4}). 

Finally dispersion relation for unstable modes takes the form
\begin{equation} \label{eq:disp2}
\int_{0}^{r_0} \frac{\sqrt{4\pi G \rho_0(r) - |\omega_n|^2}}{c_s(r)}\,dr =
\frac{\pi}{4} + n \pi, \quad n = 1,2,3 \ldots \,.
\end{equation}

Simple criterion of instability immidiently follows from expression
(\ref{eq:disp2}). If we neglect $|\omega_n|^2$ and expand integration
from $r_0$ to $R$ then we obviously get for $n = 1$ condition for existance
of instability in the form
\begin{equation} \label{eq:crit1}
\alpha = \frac{4}{5\pi} \int_0^R 
\sqrt{\frac{4\pi G \rho_0(r)}{c_s^2(r)}} \,dr > 1 \,.
\end{equation}
\section{conclusion}\label{S:frem}
Now we can apply criterion (\ref{eq:crit1}) to models of massive gaseous 
clouds mainly consisting of molecular hydrogen. At temperatures 
$\lesssim 90 \text{K}^{\circ}$ rotational degrees of freedom are 
degenerated (\cite{lL64}) and it behaves as monoatomic ideal gas with 
adiabatic exponent $\gamma = 5/3$ and polytrope exponent 
$n = 3/2$. At temperatures $>90 \text{K}^{\circ}$
adiabatic exponent has standard value $\gamma = 7/5$ and $n = 5/2$.
Criterion (\ref{eq:crit1}) may be rewritten in terms of polytrope
model as follows (cf. \ref{eq:poly1}, \ref{eq:xi1})
\begin{equation} \label{eq:crit2}
\alpha_n = \frac{4}{5\pi} \xi_n \sqrt{n}\int_0^1 
\sqrt{\Theta_n^{n-1}(\xi)} \,d\xi \,.
\end{equation}
Numeric integration in expression (\ref{eq:crit2}) 
gives for $\gamma = 5/3$
\begin{equation} \label{eq:crit3}
\alpha_{3/2} = 0.92 \,,
\end{equation}
and for $\gamma = 7/5$
\begin{equation} \label{eq:crit4}
\alpha_{5/2} = 1.1 \,.
\end{equation}
This result seems to be very intristing. It follows from
equations (\ref{eq:crit3}) and (\ref{eq:crit4}) that cold
hydrogen cloud may be stable relatively gravitational instability,
but if the mean temperature of hydrogen cloud is high enough, it
may be the subject of gravitational instability.

We can also remark here that our result for $\gamma = 5/3$ is
applicable to degenerated electron gas (white dwarfs) and
our proof of its stability is in accordance with well known
results of the theory of white dwarfs (\cite{jZ81},\cite{sW75},\cite{lL64}).


\begin{thebibliography}{99}

   \bibitem{jZ81}
	Zeldovich J.B., Blinnikov S.I., Shakura N.I. \emph{Physical base
	of structure and evolution of stars.} MSU,  Moscow, 1981, 150 p.
	(Russian)

   \bibitem{sW75}
	Weinberg S. \emph{Gravitation and cosmology.} -- Mir, 
	Moscow, 1975, 696 p. (Russian)

   \bibitem{mK99}
	Kiessling M.K. \emph{Mathematical vindication of the ``Jeans
	swindle''.} -- arXiv:astro-ph/9910247.	

   \bibitem{pChK01}
	Chavanis P.H. \emph{Gravitational instability of finite
	isothermal spheres.} -- arXiv:astro-ph/0103159.	

   \bibitem{jL01}
	Lima J.A.S, Silva R., Santos J., 
	A\&A {\bf 396}, 309 (2002). astro-ph/0109474.	

   \bibitem{aS04}
	Sandoval-Villalbazo A., Garcia-Colin L.S. \emph{Jeans instability in
	the linearized Burnett regime.} -- arXiv:astro-ph/0403249.	

   \bibitem{lL64}
      Landau L.D., Lifchitz E.M. \emph{Statistical Physics,} Nauka, Moscow, 
      1964, 567 p. (Russian).

   \bibitem{lL74}
	Landau L.D., Lifchitz E.M. \emph{Quantum mechanics.}
	Nauka,  Moscow, 1974, 752 p. (Russian).

   \bibitem{lL88}
	Landau L.D., Lifchitz E.M. \emph{Hydrodynamics.}
	Nauka,  Moscow, 1988, 733 p. (Russian).

   \bibitem{lR02}
	Rezzolla L. \emph{Gravitational Waves from Perturbed
	Black Holes and Relativistic Stars.} -- arXiv:gr-qc/0302025.

\end{thebibliography}
\end{document}